
%


 \magnification\magstep 1
 \vsize=8.5truein
 \hsize=6truein
 \voffset=0.5 truecm
 \hoffset=1truecm
 \baselineskip=22pt


\def\hef{$^4\!$He }

\def\rr{ {\bf r} }

\def\qq{ {\bf q} }
\def\jj{ {\bf J_q} }

\def\rhocq{ \rho_{\bf q}^\dagger}
\def\rhoq{ \rho_{\bf q} }
\def\omph{ \omega_\circ }
\def\jcx { J_{x{\bf q}}^\dagger }
\def\jx {J_{x{\bf q}} }
\def\jcz { J_{z{\bf q}}^\dagger }
\def\jz {J_{z{\bf q}} }

\vskip 1.5 truecm

\centerline{\bf STATIC RESPONSE FUNCTION FOR LONGITUDINAL AND}
\centerline{\bf TRANSVERSE EXCITATIONS IN SUPERFLUID HELIUM }

\vskip 2 truecm

\centerline{F. Dalfovo and S. Stringari}

\centerline{\it Dipartimento di Fisica, Universit\`a di Trento, and INFN,
I-38050 Povo, Italy}

\vskip 2.5truecm

\noindent {\bf Abstract.}\ \ {\it  The sum rule formalism is used to
evaluate rigorous  bounds for the density and current  static
response functions in  superfluid \hef at zero temperature. Both
lower and upper bounds are considered. The bounds are expressed
in terms of ground state properties (density and current correlation
functions) and of the interatomic potential. The results
for the density static response significantly improve the Feynman
approximation and turn out to be close to the experimental (neutron
scattering) data. A quantitative prediction for  the transverse
current response is given. The role of one-phonon and multi-particle
excitations in the longitudinal and transverse channels is  discussed.}

\vskip 1.5 truecm


\vfill\eject

\centerline{\bf I. INTRODUCTION}

\bigskip

It is well known that the linear response of a many body system
to an external probe (density, current,\dots) contains relevant
information on the dynamic correlations among particles [1]. If
the probe is static and coupled to the density of the system, one
has to deal with the static density response function $\chi(q)$,
related to the dynamic structure function $S(q,\omega)$ through
the equation
$$
\chi(q)  =  - 2 \int_0^\infty {S(q,\omega) \over \hbar \omega }
d\omega  \ \ \ . \eqno (1)
$$
In the low-$q$ limit this quantity is fixed by the well known
compressibility sum rule
$$
- \lim_{q \to 0} \chi(q) = {N \over M c^2} \ \ \ ,
$$
where $N$, $M$ and $c$ are the particle number, particle mass
and sound velocity respectively. In the opposite case,
$q \to \infty$, one finds the free-particle limit
$$
- \lim_{q  \to \infty} \chi(q) = {4 N M \over \hbar^2q^2} \ \ \  .
$$
In superfluid \hef the integral
of Eq.(1) can be extracted, with rather good accuracy,  from neutron
scattering experiments [2].  An important characteristic of
$\chi(q)$   is a pronounced peak in the region
of the roton wave vectors. It reflects the strong interaction
between particles which tends to produce solid-like correlations
in the system. This feature of the static response function  plays
an important role in the context of density functional
theories for inhomogeneous Bose systems [3,4].

The calculation of $\chi(q)$ represents a challenging theoretical
problem. In this work we use the sum rule formalism
to provide rigorous lower and upper bounds to the
static response function at zero  temperature. The basic
idea was already introduced by Hall and Feenberg [5], who obtained
bounds to $\chi(q)$ using the so called Feynman approximation.
The difference between the lower and the upper bounds  was, however,
too large to make their approach of practical use. More recently
an improved lower bound has been proposed [6] with the help of
additional sum rules. In the first part of this work we will
derive and evaluate explicitly new lower and upper bounds
for the static density response  function.
The ground state properties which enter the relevant sum rules,
needed to calculate the bounds, are taken from Monte Carlo
calculations [7,8].
The resulting difference between the new lower and upper bounds turns
out to be relatively small and, thus, they allow for a rather
precise estimate of the true static response in very good
agreement with the available experimental results at zero pressure.

In the second part of this work we apply the same formalism to
investigate the response to current excitations. We give a
quantitative estimate of the static response
in the transverse channel, through the use of a new lower
bound, and we compare the results with the  longitudinal case.

\vskip 2 truecm

\centerline{\bf II. \ DENSITY-DENSITY RESPONSE FUNCTION}

\bigskip

{\bf 2.1 \ General formalism}

\smallskip

The linear response function characterizes the behavior of a quantum
many body system subject to a small external perturbation.  For static
density excitations, at zero
temperature,  one can write the total Hamiltonian as
$$
H(\lambda) = H + \lambda \rhocq  \ \ \ , \eqno (2)
$$
where
$$
H= \sum_j {p_j^2 \over 2 M} + \sum_{i\neq j} V(|\rr_i-\rr_j|) \eqno(3)
$$
is the unperturbed Hamiltonian, $\lambda$ is the strength of
the perturbing field, and
$$
\rhocq = \sum_{j=1}^N e^{i\qq \cdot \rr_j} \eqno (4)
$$
is the usual density operator. The static response function is
defined as
$$
\chi(q) = \lim_{\lambda \to 0} {\langle \lambda | \rhoq | \lambda
\rangle \over \lambda } \ \ \ , \eqno (5)
$$
where $|\lambda \rangle$ is the ground state of $H(\lambda)$. At
zero temperature, standard perturbation theory gives $\chi(q)$
in terms of the eigenstates and eigenvalues of $H$ as follows:
$$
\chi(q)= -2 \sum_n { |\langle n |\rhocq |0\rangle |^2 \over \hbar
\omega_{n0} }  \ \ \ . \eqno (6)
$$
With the usual definition of the dynamic structure function
$$
S(q, \omega)  =   \sum_n | \langle n | \rhocq | 0 \rangle |^2 \delta
(\omega -\omega_{n0})   \eqno (7)
$$
and its moments
$$
m_p(q) = \int_0^\infty (\hbar \omega)^p S(q,\omega) d\omega
= \sum_n |\langle n | \rhocq |0\rangle |^2 (\hbar \omega_{n0})^p
\ \  \ ,
\eqno (8)
$$
Eq.(6) reads
$$
\chi(q)  =  - 2 \int_0^\infty {S(q,\omega) \over \hbar \omega }
d\omega = - 2 m_{-1} (q) \ \ \ , \eqno (9)
$$
The dynamic structure function  $S(q,\omega)$ contains the detailed
structure of the elementary excitations of the system. It is
directly accessible to inelastic neutron scattering experiments on a
relevant range of wave vectors. In particular, the integral in Eq.(9)
can be estimated rather precisely from the observed spectrum; the
 factor $1/\omega$ makes the integral rapidly convergent at high
$\omega$, reducing the effects of the complex structure of multiphonon
excitations. An accurate measurement of $\chi(q)$ is available, at
present, only at zero pressure [2]. The theoretical determination
of $\chi(q)$ is a much harder problem; a direct approach, in fact,
implies the calculation of non-uniform perturbed states with high
enough accuracy to extract the linear limit; first results on this
line are now becoming available [9].
In this
work we choose an alternative approach based on the use of the moments
$m_p$ with  $p \ge 0$, which, differently from $m_{-1}$,  can be
expressed in terms of known ground state properties with the help
of sum rules.

By using the completeness relationship $\sum_n |n\rangle \langle
n|= {\bf 1}$  in Eq.(8),  one finds
$$
m_0(q) = \int \! d\omega \ S(q,\omega) = \langle \rho_\qq^\dagger
\rho_\qq \rangle \eqno (10a)
$$
$$
m_1(q) = \int \! d\omega \ \hbar \omega \ S(q,\omega) = {1 \over 2 }
 \langle [\rho_\qq^\dagger,[H,\rho_\qq]] \rangle \eqno(10b)
$$
$$
m_2(q) = \int \! d\omega \ (\hbar \omega)^2 \ S(q,\omega) =
\langle [\rho_\qq^\dagger,H][H,\rho_\qq] \rangle \eqno (10c)
$$
$$
m_3(q) = \int \! d\omega \ (\hbar \omega)^3 \ S(q,\omega) = {1 \over 2 }
 \langle [[\rho_\qq^\dagger,H],[H,[H,\rho_\qq]]] \rangle \ \ \ .
\eqno (10d)
$$
The above equations relate properties of the excitation spectrum to mean
values on  the ground state. They are known as sum rules and have been
extensively used in the theory of Bose  liquids [1,5,6].
The moment $m_0$ coincides  with the static form factor
$$
N S(q) =  m_0(q)   \ \ \ , \eqno (11)
$$
where $N$ is the number of particles. $S(q)$ is related to
the radial distribution function $g(r)$ by means of
$$
S(q) -1 = \int \! d\rr \ (g(r)-1) \ e^{i \qq \cdot \rr} \ \ \ .
\eqno (12)
$$
Several {\it ab initio} calculations are available for the pair
correlation function $g(r)$ (see for example Refs.~7 and 8). The
corresponding $S(q)$ well agrees with the experimental static form
factor [10].

The energy weighted sum rule $m_1$ is the model independent f-sum rule
$$
m_1 (q) = N { \hbar^2 q^2  \over 2M} \ \ \ , \eqno (13)
$$
which follows from the particle number conservation [1]. It can be easily
derived from Eq.(10b) taking into account that the interatomic potential
commutes with the density operator.

The $m_2$ sum rule can be expressed in terms of the current correlation
function using the continuity equation
$$
[\rhoq, H] = \hbar \qq \cdot \jj \ \ , \eqno (14)
$$
with the current density operator given by
$$
\jj = {1\over2} \sum_{j=1}^N \left( {{\bf p}_j \over M} e^{-i \qq \cdot
\rr_j} + e^{-i \qq \cdot \rr_j} {{\bf p}_j \over M} \right) \ \ .
\eqno (15)
$$
If  $\qq$ is taken along  $z$,  one finds
$$
m_2 (q)  = \hbar^2 q^2  \langle \jcz \jz \rangle
\ \ \ . \eqno(16)
$$
Eq.(16) shows that $m_2(q)$ is proportional to the longitudinal
component of the current correlation function in $\qq$-space.
The  $m_2$ moment can be also written in the following form [5]:
$$
m_2 (q) = N \left[ \left( 2-S(q) \right) \left( { \hbar^2 q^2 \over 2 M}
\right)^2 + {\hbar ^4 q^2 \over M^2} D(q) \right]  \ \ \ , \eqno (17)
$$
where $D(q)$ is the so called kinetic structure function
$$
D(q) =  \int d \rr_1 d \rr_2 \cos( q(z_1-z_2))
 \nabla^z_1  \nabla^z_2
\rho^{(2)}( \rr_1, \rr_2;  \rr_1^{\prime},  \rr_2^{\prime})\mid
_{ \rr_1=  \rr_1^{\prime},  \rr_2 =  \rr_2^{\prime}} \ \ .
\eqno(18)
$$
While $S(q)$ is fixed  by the diagonal components of the two-body
density matrix, the kinetic structure function $D(q)$ requires the
knowledge of the non-diagonal components. At $P=0$ we can use the
Path Integral Monte Carlo
calculations of Pollock and Ceperley [8,11] for the current density
correlations in order to evaluate $D(q)$. In Fig.~1 we plot the resulting
curve. At $q$ smaller than about $1$ \AA$^{-1}$ the accuracy is poor,
due to the finite-size box of the PIMC calculations. At large $q$
the numerical results are consistent with the asymptotic limit [5]
$$
\lim_{q \to \infty} D(q) =  {2 \over 3} {M \over \hbar^2} \langle
\hbox{KE} \rangle  \ \ \ ,  \eqno (19)
$$
where $\langle \hbox{KE} \rangle$ is the mean kinetic energy
per particle in the ground state.

Finally we note that the
cubic energy weighted moment $m_3$ can be rather easily evaluated
carrying out the commutators in Eq.(10d). One finds [12]
$$
m_3 (q) = N \! \left[ ({\hbar^2 q^2 \over 2 M})^3 + {\hbar^4 q^4 \over M^2}
\langle \hbox{KE} \rangle + { \rho_\circ \hbar^4 \over 2 M^2}
\int \! \! d\rr \ g(r) (1 \! - \! \cos(\qq \! \cdot \! \rr)) (\qq \! \cdot \!
\nabla)^2 V(r) \right]
\ , \eqno (20)
$$
where  $\rho_\circ$ and  $V(r)$ are the particle
density and the interatomic potential respectively.

\bigskip
\bigskip

{\bf  2.2  \ The Feynman approximation}

\smallskip

So far we have shown how the sum rules $m_0$, $m_1$, $m_2$ and $m_3$
can be determined from known properties of the ground state. Now
we use them to fix rigorous bounds to $m_{-1}$ and, consequently,
to the static response function $\chi(q)$. We notice that $S(q,\omega)$
is  a positive function and  the inequality
$$
\int_0^{\infty} d\omega {S(q,\omega) \over \hbar \omega}
(1+ \alpha \hbar \omega)^2  \ge 0  \eqno (21)
$$
holds for any real $\alpha$. Using the definition of the moments $m_p$,
one has
$$
m_{-1} \ge - (2\alpha m_0 + \alpha^2 m_1 ) \ \ \ . \eqno (22)
$$
One can vary the parameter $\alpha $ to make the r.h.s. of eq.(22)
maximum. This yields $\alpha = - m_0/m_1$ and
$$
m_{-1} \ge {(m_0)^2 \over m_1} = {N S(q) \over \hbar \omega_{_F}(q)} \ \ \ ,
\eqno (23)
$$
where
$$
\hbar \omega_{_F} (q) = {m_1(q) \over m_0(q)} \eqno (24)
$$
is the energy of the phonon-roton excitation branch in the so called
Feynman  approximation. Equation (23) provides a first rigorous bound to the
static response function at $T=0$.
Using Eqs. (11) and (13) the same inequality
can be written in the form [5]
$$
m_{-1}(q)  \ge m_{-1}^{F}(q) = {2 N M S^2(q) \over \hbar^2 q^2} \ \ \ .
\eqno (25)
$$
The quantity $m_{-1}^{F}(q)$ corresponds to the Feynman approximation to
$m_{-1}$.

In a similar way one can find an upper bound to $\chi(q)$.  The crucial
point is that in superfluid \hef there are no excitations with
energy lower than the  energy $\hbar \omph$ of the phonon-roton
branch. Thus, it is possible to write
$$
\int_0^\infty  {S(q,\omega) \over \hbar \omega} \ d\omega \le
\int_0^\infty { S(q,\omega)
\over \hbar \omph} \ d\omega \ \ \ , \eqno (26)
$$
which implies [5]
$$
m_{-1} \le { m_0 \over \hbar \omph } \  \ \ . \eqno (27)
$$
Precise measurements of $\hbar \omph$ are available [2,13], so
that the upper bound (27) can be accurately estimated at several
pressures.  One notices that the two bounds (25) and (27) would
collapse in  the exact $m_{-1}$ if the excitation  spectrum
were exhausted by a single collective phonon-roton mode. In that
case also the Feynman energy $\hbar \omega_{_F}$ would coincide
with the true  phonon-roton energy $\hbar \omph$.

In Fig.~2  the lower (25) and upper (27) bounds,
calculated at zero pressure, are plotted as dashed lines, together
with the experimental data. Experiments  are consistent with the
theoretical bounds which, however, turn out to be quite far
each other. The sizable difference between the two bounds is a
measure of the role of multiphonon excitations
and is a signature of the inadequacy of the Feynman approximation.

\bigskip
\bigskip

{\bf  2.3 \ New  bounds for $m_{-1}$}

\bigskip

Equations (21) and (26) can be generalized in a natural way through the
proper inclusion of additional sum rules.  Let's begin with
the inequality
$$
\int_0^\infty  d\omega {S(q,\omega) \over \hbar \omega}
(1 + \alpha \hbar \omega
+ \beta \hbar^2 \omega^2)^2 \ge 0 \  \ \ , \eqno (28)
$$
valid for any real $\alpha$ and $\beta$. As before, we can write Eq.(28)
as a lower bound for $m_{-1}$ and vary both $\alpha$
and $\beta$. After a straightforward calculation one gets [6]
$$
m_{-1}(q) \ge {m_{-1}^{F}(q) \over   1- \Delta(q)/\epsilon(q) } \ \ \ ,
\eqno (29)
$$
where
$$
\epsilon(q) = \left[ {m_3 \over m_1}+({m_1 \over m_0})^2-2{m_2 \over m_0}
\right] \left( {m_2 \over m_1}-{m_1 \over m_0} \right)^{-1}  \eqno (30)
$$
and
$$
\Delta(q) = {m_2 \over m_1}-{m_1 \over m_0} \ \ \ . \eqno(31)
$$
The ratio $\Delta/\epsilon$ takes important contributions from
multiphonon excitations, through the moments $m_2$ and $m_3$.
As a consequence we expect a significant improvement with
respect to the Feynman approximation (25).

As concerns the upper bound  we generalize Eq.(26) in the
following way [14]:
$$
\int_0^\infty d\omega {S(q,\omega) \over \hbar \omega}
(1+ \gamma \hbar \omega)^2
\le \int_0^\infty d\omega {S(q,\omega) \over \hbar \omph} (1+ \gamma \hbar
\omega)^2  \eqno (32)
$$
or, equivalently,
$$
m_{-1} \le {m_0 \over \hbar \omph} + 2 \gamma \left( {m_1 \over \hbar
\omph} - m_0 \right) + \gamma^2 \left( {m_2 \over \hbar \omph } -
m_1 \right) \ \ \ . \eqno (33)
$$
Minimization with respect to $\gamma$ yields
$$
m_{-1}(q) \le {m_0 \over \hbar \omph} \left[ 1- {m_0 \over m_1}
\left( {m_1 \over m_0} - \hbar \omph \right)^2 \left( {m_2 \over m_1}
-\hbar \omph \right)^{-1} \right] \ \ \ . \eqno (34)
$$
Note that since both $m_1/m_0$ and $m_2/m_1$ differ from $\hbar \omph$,
due to the important role of multiphonon excitations, Eq.(34) yields
a significant lowering with respect to the Feynman upper bound (27).
In Fig.~2 the new bounds (29) and (34) at zero pressure are plotted
as solid lines. At small $q$ the microscopic ingredients
used in the present sum rule analysis (density and current correlation
functions of  the ground state) suffer from a lack of accuracy
due to the finite-size of the cell for Monte Carlo simulations. For this
reason we have not shown the curves for the upper and lower bounds
below $q \simeq 1$\AA$^{-1}$. As in the case of the Feynman approximation
the experimental data fulfil the theoretical bounds. But now the
two bounds are much closer each other over all the relevant range
of $q$'s. This means that Eqs. (29) and (34) provide an estimate of
the static response function close to the exact value. We stress
again that the evaluation of the two bounds (29) and (34) involves
only ground state properties and, consequently, is much simpler
than the explicit {\it ab initio} calculation  of the static response
function.

\vskip 2 truecm

\centerline{\bf III. \ THE CURRENT-CURRENT RESPONSE}

\bigskip

{\bf  3.1 \ Longitudinal  current excitations }

\smallskip

In this section we rewrite the formalism of Section 2.1 for the current
response function. The current operator has been already defined
in Eq.(15). As before, one adds a small perturbation to the Hamiltonian
of the system. The perturbation is now a vector field proportional
to the current density operator. The response of the system is
given by the current  response tensor
$$
\chi_{\mu \nu} (q,\omega) = \sum_n \left[ { \langle 0| J_{\mu{\bf q}} |n
\rangle \langle n| J_{\nu{\bf q}}^\dagger  |0 \rangle \over
\hbar \omega - \hbar\omega_{n0} + i\eta} - { \langle 0|
J_{\nu {\bf q}}^\dagger |n \rangle \langle n | J_{\mu {\bf q}} |0 \rangle
\over \hbar \omega + \hbar \omega_{n0} +i\eta } \right]
\  \  \ . \eqno(35)
$$
The transverse and longitudinal components of the response  tensor
can be studied separately. Let's begin with the longitudinal  one. We
take $\qq$ along $z$ and define the longitudinal response function
$$
\chi^L(q,\omega) = \sum_n {2 \hbar \omega_{n0} |\langle n |\jcz |0\rangle
|^2 \over (\hbar \omega +i \eta)^2 - (\hbar\omega_{n0})^2 } \ \ \ .
\eqno(36)
$$
Then we define the quantity
$$
\Upsilon^L (q,\omega) = \sum_n |\langle n|\jcz|0\rangle|^2
\delta(\omega- \omega_{n0})           \eqno(37)
$$
and its moments
$$
m^L_p(q) = \int_0^\infty (\hbar \omega)^p \Upsilon^L(q,\omega)
 d\omega = \sum_n |\langle n|\jcz|0\rangle|^2 (\hbar \omega_{n0})^p
\ \ \ .  \eqno (38)
$$
The longitudinal static response function is the $\omega \to 0$
limit of Eq.(36). Using the definition of $\Upsilon^L$ one has
$$
- \chi^L(q) =   2 \int\! d\omega {\Upsilon^L (q,\omega) \over
\hbar \omega} \ \ \  . \eqno (39)
$$
Indeed the determination of the  static response function (39) is
trivial. The key point is the continuity equation (14) which
connects the matrix elements of the longitudinal current with
the ones of the density operator:
$$
\langle n | \jcz |0 \rangle = { \omega_{n0} \over  q}
\langle n| \rhocq |0\rangle \ \ \ . \eqno (40)
$$
One easily obtains [1,15]
$$
- \chi^L(q) = 2 m^L_{-1} (q) = 2 {m_1(q) \over \hbar^2 q^2} =
{N \over M}  \ \ \ , \eqno (41)
$$
where $m_1$ is the density f-sum rule (13). In the same way one
finds
$$
m^L_0(q) = \langle \jcz \jz \rangle = {m_2(q) \over \hbar^2 q^2}
\eqno (42)
$$
and
$$
m^L_1(q) = {1\over 2} \langle [[\jcz,H],\jz] \rangle = {m_3(q)
\over \hbar^2 q^2} \ \ \ , \eqno (43)
$$
where $m_2$ and $m_3$ are the sum rules (16) and (20).
The simplicity  of results (41)  reflects the fact that the response
to a static longitudinal probe is, actually, a fictitious problem,
related to gauge  invariance properties [1].

\bigskip
\bigskip

{\bf 3.2 \ Transverse current excitations }

\smallskip

The response to transverse probes plays a crucial role in the
theory of superfluidity. Actually the $q \to 0$ limit of the
transverse response function defines the normal (non
superfluid) density [1,15] of the system. This limit was extracted
in Ref.~8 at several temperatures through a Path Integral
Monte Carlo calculation of the transverse current correlation
function and the use of the fluctuation-dissipation theorem.
In this section we provide a first estimate of $\chi^T$
at finite $q$ by calculating a rigorous lower bound at
zero temperature.

Let's rewrite Eqs. (36-39) by replacing the superscript $L$
with $T$ and the $z$-component of the current with an arbitrary
component orthogonal to $\qq$. Similarly to the longitudinal case
one has
$$
m^T_0 (q) = \langle \jcx \jx \rangle \eqno(44)
$$
and
$$
m^T_1 (q) = {1\over 2} \langle [[\jcx ,H],\jx] \rangle \ \ \ .
\eqno (45)
$$
The calculation of these two sum rules follows exactly the
procedure used for the $m_2$ and $m_3$ sum rules in the case
of density excitations (see Eqs. (16) and (20)).
In particular, the moment $m^T_0$ is the  Fourier transform
of the transverse current  correlation function.
The $m_1^T$ moment can be evaluated by
carrying out explicitly the commutators in Eq.(45). One finds
$$
m_1^T(q) = N \left[ {\hbar^2 q^2 \over 3 M^2} \langle KE \rangle +
{\rho_\circ \hbar^2 \over 2 M^2}  \int\! d\rr \ g(r) (1-\cos(qz))
\nabla^2_x V(r) \right]
\ \ \  .  \eqno(46)
$$
The transverse static response function is given by
$$
- \chi^T (q) = 2 m^T_{-1} = 2 \int d\omega {\Upsilon^T (q,\omega) \over
\hbar \omega} \ \ \ . \eqno (47)
$$
The quantity $\Upsilon^T(q,\omega)$ is  positive, so that
the inequality (22) holds even for the transverse moments
and, thus,
$$
- \chi^T(q) \ge 2 {(m^T_0(q))^2 \over m^T_1(q)} \ \ \ . \eqno(48)
$$
This is a rigorous lower bound, valid at zero temperature.
We have explicitly evaluated the r.h.s. of Eq.(48) at zero temperature,
taking  $m^T_0$ from the transverse current
correlation function calculated in Ref.~8.
The resulting lower bound is shown in Fig.~3.  Since
the PIMC  calculations  of Ref.~8 are  carried out at low but
finite temperature (the lowest value is $T=1.18$ K), the use of  the
corresponding $m^T_0$ in Eq.(48) is meaningful  only for $q$ much
greater than $kT/c$; the curve in Fig.~3 corresponds to values
of $q$ well above this limit.

At zero temperature  the function $\chi^T(q)$
should vanish at $q=0$ because the
transverse current operator cannot excite phonons, which, in this
limit,  are the  dominant excitations   (the
system is entirely superfluid).
At higher $q$ multiphonon processes take place
and $\chi^T(q)$ no longer vanishes. The first contribution
to the static response is expected to be in $q^2$.
In the opposite case $q \to \infty$ one approaches the
free particle limit and  all the ratios $m_{p+1}/m_p$ tend
to the  same energy $\hbar^2 q^2/2M$. This implies
$$
- \lim_{q \to \infty} \chi^T (q) = 2 \lim_{q \to \infty} {(m^T_0 (q))^2
\over m^T_1 } = N {8 \over 3} {\langle KE \rangle \over \hbar^2 q^2}
\ \ \ , \eqno (49)
$$
where we have used the asymptotic values
$$
\lim_{q \to \infty} m^T_0 = {2 N \over 3 M} \langle KE \rangle
\eqno (50)
$$
and
$$
\lim_{q \to \infty} m^T_1 = {N \hbar^2 q^2 \over 3 M^2} \langle KE
\rangle \ \ \ . \eqno (51)
$$
The asymptotic behavior of $\chi^T(q)$ is shown in Fig.~3 as a
dot-dashed line. The position of the maximum of the solid curve
provides  a characteristic coherence length for superfluidity [1].
As expected, it is of the same order as the
roton wave vector. The height at the maximum measures the strength of
the interaction between particles. In fact in a free Bose gas at zero
temperature the function $\chi^T(q)$ would be identically zero.

Clearly Eq.(48) provides only a lower bound for $-\chi^T(q)$.
A rough estimate of the difference between this lower bound and the
exact value of the transverse static response can be made using the
following  arguments. First we note that only
multiparticle excitations affect the transverse response
function, since no transverse current is carried by the elementary
excitations (phonons, rotons) of the system. If one assumes
that the average energy and the spreading of multiparticle excitations
are the same in the longitudinal as in the transverse channels,
one can estimate the relative difference between $m^T_{-1}$ and
$(m^T_0)^2/m^T_1$ by analysing the multiparticle contribution to
the corresponding longitudinal sum rules. Such a contribution
can be explicitly extracted from the experimental data on
$S(q,\omega)$ [2]. With this procedure we conclude that the lower
bound (48) should underestimate the exact value of $- \chi^T(q)$ by
about $30$\% in the roton region.

\vskip 2 truecm

\centerline{\bf IV. \ CONCLUSIONS}

\bigskip

In this work we have investigated lower and upper bounds for the
static response function of superfluid \hef at zero temperature. In
the case of the density response $\chi(q)$  the
new bounds improve significantly the Feynman approximation,
yielding estimates in agreement with the experimental data.
In the case of the current response function we have given a
first rigorous bound to the static response $\chi^T(q)$ and made a
direct comparison with the longitudinal response, stressing the
different role of one-phonon and multiparticle excitations.

\vskip 2 truecm

{\bf ACKNOWLEDGEMENTS}

\smallskip

We thank  D.M.  Ceperley and E.L. Pollock for providing
us with their numerical data on current density  correlations.
Useful discussions with R.M. Bowley and L. Pitaevskii are
acknowledged.

\vfill\eject

\centerline{\bf REFERENCES}

\bigskip
\bigskip

\item{[1] } D. Pines and P. Nozi\`eres, {\it The Theory of Quantum
Liquids} (Benjamin, New York 1966), Vol.I;
 P. Nozi\`eres and D. Pines, {\it The Theory of Quantum Liquids},
(Addison Wesley, 1990), Vol.II.

\item{[2] } R.A. Cowley and A.D.B. Woods, Can. J. Phys. {\bf 49}, 177
(1971); A.D.B Woods and R.A. Cowley, Rep. Prog. Phys. {\bf 36}, 1135
(1973).

\item{[3] } J. Dupont-Roc, M. Himbert, N. Pavloff and J. Treiner, J. Low
Temp. Phys. {\bf 81}, 31 (1990).

\item{[4] } F. Dalfovo, J. Dupont-Roc, N. Pavloff, S.Stringari and J.
Treiner, Europhys. Lett. {\bf 16}, 205 (1991); S. Moroni and G. Senatore,
{\it ibid.} {\bf 16}, 373 (1991).

\item{[5] } E.Feenberg, {\it Theory of Quantum Fluids}, (Academic Press,
New York and London, 1969) ch.4 ; D.Hall and E.Feenberg, Ann. of Phys.
{\bf 63}, 335 (1971).

\item{[6] } S. Stringari, Phys. Rev. {\bf B}, in press.

\item{[7] } M. Kalos, M.A. Lee, P.A. Whitlock and G.V. Chester, Phys.
Rev. {\bf B 24}, 115 (1981).

\item{[8] } E.L. Pollock and D.M. Ceperley, Phys. Rev. {\bf B 36}, 8343
(1987).

\item{[9] } D.M. Ceperley ,S.Moroni and G.Senatore, private communication.

\item{[10] } E.C. Svensson, V.F. Sears, A.D.B. Woods and P. Martel,
Phys. Rev. {\bf 21}, 3638 (1980).

\item{[11] } In the calculations of Ref.~8 a source of inaccuracy for
the current-current correlation function, which enter our $m_2$ sum rule,
is the so called end-point approximation. One consequence is that the
value of the mean kinetic energy per particle is overestimated by
about $1$K. We checked that this problem has negligible quantitative
effects on our results.

\item{[12] } R.D. Puff, Phys. Rev. {\bf A 137}, 406 (1965).

\item{[13] } R.J. Donnelly, J.A. Donnelly and R.N. Hills, J. Low
Temp. Phys. {\bf 44}, 471 (1981).

\item{[14] } The idea of such a generalization was suggested to us
by R.M. Bowley.

\item{[15] } G. Baym, in {\it Mathematical Methods in Solid State and
Superfluid Theory}, R.C. Clark and G.H. Derrick eds. (Oliver and Boyd,
Edinburgh, 1969) p.121.

\vfill \eject

\centerline{\bf FIGURE CAPTIONS}

\vskip 2 truecm

\item{Fig.~1 }  Kinetic structure function extracted from Eq.(17)
               with $m_2(q)$ and $S(q)$ taken from Monte Carlo
               calculations [7,8]. In the limit $q \to \infty$ one finds
               the asymptotic value $0.8$ \AA$^{-2}$, as in Eq.(19).

\item{Fig.~2 }  Static response function for density excitations. Open
               circles: experimental values [2]; dashed lines: upper
               and lower bounds in Feynman approximation; solid lines:
               upper and lower bounds given in Section 2.3.

\item{Fig.~3 }  Static response function for transverse current
               excitations. Solid line: lower bound (48); dot-dashed
               line: $q \to \infty$ asymptotic curve for the exact
               response function (see Eq.(49)).

\bye